\documentclass[11pt]{amsart}

\usepackage[utf8]{inputenc}
\usepackage{amssymb}
\usepackage{epsfig}
\usepackage{comment}
\usepackage{amsmath}
\usepackage[section]{placeins}
\usepackage{mathrsfs}
\usepackage{graphicx}
\usepackage[notrig]{physics}
\usepackage{subfigure}
\usepackage{caption}
\usepackage{color}
\usepackage{svg}
\usepackage{Lib-Notation} 
\usepackage{siunitx}

\graphicspath
{
  {./}
  {./Figures/}
}

\begin{document}

\title{Thermalized and mixed meanfield ADP potentials for magnesium hydrides}
\author{
M.~Molinos${}^1$, M.~Ortiz${}^{2,3}$ and M.~P.~Ariza${}^1$
}

\address
{
  ${}^1$Escuela T\'ecnica Superior de Ingenier\'ia, Universidad de Sevilla, Camino de los descubrimientos, s.n., 41092 Sevilla, Spain.
  \\
  ${}^2$Division of Engineering and Applied Science, California Institute of Technology, 1200 E.~California Blvd., Pasadena, CA 91125, USA.
  \\
  ${}^3$Institut f\"ur Angewandte Mathematik and Hausdorff Center for Mathematics, Universit\"at Bonn, Endenicher Allee 60, 53115 Bonn, Germany.
 }

\email{$\dots$}

\begin{abstract}
We develop meanfield approximation and numerical quadrature schemes for the evaluation of Angular-Dependent interatomic Potentials (ADPs) for magnesium and magnesium hydrides at finite temperature (thermalization) and arbitrary atomic molar fractions (mixing) within a non-equilibrium statistical mechanical framework and derive local equilibrium relations.  We numerically verify and experimentally validate the accuracy and fidelity of the resulting thermalized/mixed ADPs (TADPs) by means of selected numerical tests including free entropy, heat capacity, thermal expansion, molar volumes, equation of state and elastic constants. We show that the local equilibrium properties predicted by TADPs agree closely with those computed directly from ADP by means of Molecular Dynamics (MD).
\end{abstract}

\maketitle

\section{Introduction}
\label{sec:intro}

Hydrogen energy storage is a challenging technological field that attracts the efforts of a broad research community \cite{Yang_2021} and elicits the deployment of a variety of tools from SEM and TEM \cite{Li2018} to DFT simulations \cite{Tao,KLYUKIN2015371} and {\sl ab initio} \cite{KLYUKIN2013S10,Junkaew_et_al_2014}. The complexities inherent to magnesium-based materials are well-known \cite{Tan2021}. These complexities are amplified in the presence of hydrogen diffusion, which is controlled by the kinetics of the magnesium hydride phases. Several studies have been devoted to gaining an understanding of the rate-controlling mechanisms attendant to hydrogen diffusion in magnesium \cite{LUO201958}, including size and shell effects \cite{Shriniwasan, Pundt, Shen}.

A persistent fundamental difficulty that arises in the study of atomic diffusion concerns the staggering gaps in time scales between the atomic-scale rate-controlling mechanisms and the macroscopic behavior that they engender. Thus, direct molecular dynamics (MD) requires the use of time steps of the order of the period of thermal vibration of the atoms and, therefore, is ill-suited to the study slow long-term phenomena. Likewise, Monte Carlo methods require the enumeration of transition paths and the elucidation of the corresponding transition rates, which adds a level of empiricism on top of molecular dynamics and renders the approach intractable when the transition paths are numerous and complex, e.~g., such as they are across phase and grain boundaries.

These issues have motivated the development of analysis techniques capable of retaining full atomistic resolution while simultaneously enabling the simulation of long-term, diffusion mediated, non-equilibrium transport phenomena, including heat conduction \cite{Kulkarni2006, kulkarni2008, ArizaQC, Ponga:2015, Ponga:2016} and mass transport \cite{Venturini2011, Li2011, Sarkar2012, Venturini2014, Simpson2016}. Following the terminology of \cite{Li2011}, we collectively refer to these techniques as Diffusive Molecular Dynamics (DMD).

The essence of the approach is to describe the state of a system as a collection of {\sl sites} that can be occupied or empty. A statistical-mechanical treatment of the ensemble then allows the sites to be partially occupied as well. The degree of occupancy of a site is measured by a local atomic molar fraction variable ranging from $0$ (empty) to $1$ (fully occupied). We specifically employ the non-equilibrium statistical mechanics framework developed by \cite{Kulkarni2006, kulkarni2008} and extended by \cite{Venturini2011, Venturini2014}.

The evolution of the local atomic molar fraction variables is governed by an atomic-level kinetic equation akin to Ficks law and in the spirit of Onsager kinetics. In this framework, the atomic-level kinetics is controlled by an empirical master equation that governs the evolution of the atomic molar fractions at each individual site \cite{Li2011}. The driving forces for this evolution derive from non-equilibrium atomic-level chemical potentials. These chemical potentials in turn follow directly from a free-entropy function, defined in terms of standard interatomic potentials, which generalizes the classical Helmholtz free energy to non-equi\-lib\-ri\-um conditions. Thus, the time evolution of the system may be viewed as the result of a competition between free-entropy maximization and atomic-level kinetics. The DMD framework has been successfully applied to thermo\-mechanical dynamic problems \cite{Kulkarni2006, kulkarni2008, ArizaQC, Ponga:2015, Ponga:2016, Ponga:2017}, including heat conduction, and mass transport problems \cite{Venturini2011, Venturini2014, Wang:2015, Sun:2017}.

A first step in the deployment of DMD for specific materials systems is to characterize their local equilibrium properties at finite temperature.
In the present work, we focus on this aspect of DMD for magnesium and its hydrides, as characterized by the Angular-Dependent interatomic Potential (ADP) proposed by \cite{SMIRNOVA}. ADPs were originally developed by \cite{MISHIN2006} in order to improve the fidelity of Embedded Atom Method potentials (EAM) and modified EAM (MEAM) potentials \cite{MEAM}. Following \cite{Kulkarni2006, kulkarni2008, Venturini2011, Venturini2014}, we evaluate the free-entropy of magnesium hydrides at finite temperature by a combination of computationally efficient approximation techniques, including meanfield approximation and numerical quadrature. The approach generates {\sl thermalized} and {\sl mixed} ADPs (TADPs) for magnesium hydride phases on-the-fly that can be used as part of DMD calculations. We verify the accuracy of TADPs by means of selected numerical tests and we validate the thermalized potentials against experimental data, including free entropy, heat capacity, thermal expansion, molar volumes, equation of state and elastic constants. This verification and validation (V\&V) study reveals an overall good numerical accuracy and physical fidelity of TADPs, and establishes thermalization and mixing as a robust and efficient alternative to MD.

\section{Theoretical basis}

By way of background and to set the framework, we summarize salient aspects of the non-equilibrium statistical mechanics theory of \cite{Venturini2014}.

\subsection{Non-equilibrium statistical mechanics}

We consider a system consisting of $N$ {\sl sites}, e.~g., atomic positions or molecules, each of which can be of one of $M$ species. For each site $i=1,\dots,N$, and each species $k=1,\dots,M$, we introduce the {\sl occupancy function}
\begin{equation}\label{eq:ESM:x}
    n_{ik}
    =
    \left\{
    \begin{array}{ll}
    1, & \text{if site } i \text{ is occupied by species } k, \\
    0, & \text{otherwise},
    \end{array}
    \right.
\end{equation}
in order to describe the {\sl occupancy} of each site. We note that, from
definition~\eqref{eq:ESM:x}, we must have
\begin{equation}\label{eq:ESM:xnorm}
    \sum_{k=1}^M n_{ik} = 1
\end{equation}
at every site $i$. We additionally denote by ${n}_i = (n_{ik})_{k=1}^M$ the local occupancy array of site $i$. The {\sl microscopic states} of the system are defined by the instantaneous position $\{{q}\} = ({q}_{i})_{i=1}^{N}$, momenta $\{{p}\} = ({p}_{i})_{i=1}^{N}$, and occupancy arrays $\{{n}\} = ({n}_{i})_{i=1}^{N}$ of all $N$ sites in the system. The occupancy functions~${n}_i$ take values in a set $\mathcal{O}_M$ consisting of the elements of~$\{0,1\}^M$ that satisfy the constraint~\eqref{eq:ESM:xnorm}. In addition, the occupancy arrays $\{{n}\}$ take values in the set
\begin{equation} \label{eq:ESM:C}
\begin{split}
    &
   \mathcal{O}_{NM}
   =
   \Big\{
       \{{n}\}\in\{0,1\}^{NM}
       \, : \, \\ & \qquad\qquad\qquad
       {n}_i \in \mathcal{O}_M
       \ \mathrm{for}\
       i=1,\ldots,N
   \Big\}.
\end{split}
\end{equation}
The expected or {\sl macroscopic value} of a function $A(\{{q}\},\\ \{{p}\},\{{n}\})$ is given by the {\sl phase average}
\begin{equation}\label{eq:ESM:av}
\begin{split}
    &\langle A \rangle
    =
    \sum_{\{{n}\}\in\mathcal{O}_{NM}}
    \frac{1}{h^{3N}} \times \\ & \qquad
    \int_{\Gamma}
    A(\{{q}\}, \{{p}\}, \{{n}\})
    \;
    \rho(\{{q}\}, \{{p}\}, \{{n}\})
    \, dq \, dp ,
\end{split}
\end{equation}
where $\Gamma=(\mathbb{R}^3\times\mathbb{R}^3)^N$, $h$ is Planck's constant and $h^{-3N}$ supplies the natural unit of phase volume for systems of {\sl distinguishable particles} \cite{Hill-87, Girifalco-00}.

We assume that the statistics of the system obeys Jaynes' {\sl principle of maximum entropy} \cite{Jaynes1957a, Jaynes1957b, Zubarev, Callen-85}, which posits that the probability density function $\rho(\{{q}\}, \{{p}\}, \{{n}\})$, characterizing the probability of finding the system in a state $(\{{q}\}, \{{p}\}, \{{n}\})$, maximizes the {\sl information-theoretical entropy}
\begin{equation}\label{eq:ESM:Sgeneral}
    \mathcal{S}[\rho] = - k_B \langle \log \rho \rangle ,
\end{equation}
among all probability measures consistent with the constraints on the system. In (\ref{eq:ESM:Sgeneral}) and subsequently, $k_B$ denotes Bolzmann's constant.

We consider systems consisting of distinguishable particles whose Hamiltonians have the additive structure
\begin{equation}\label{eq:NSM:hi}
    H(\{{q}\}, \{{p}\}, \{{n}\}) = \sum_{i=1}^N h_i(\{{q}\}, \{{p}\}, \{{n}\}) ,
\end{equation}
where $h_i$ is the {\sl local Hamiltonian} of particle $i$. Suppose that the expected {\sl particle energies} and the expected {\sl particle atomic fractions}
\begin{equation}\label{eq:NSM:E}
    \langle h_i \rangle
    =
    e_i ,
    \qquad
    \langle n_{ik} \rangle
    =
    \chi_{ik},
\end{equation}
respectively, are known. We note that the local atomic fractions satisfy the identities
\begin{equation}\label{eq:NSM:N}
    \sum_{k=1}^M \chi_{ik}
    =
    1 ,
\end{equation}
which follows from (\ref{eq:ESM:xnorm}) and the second of (\ref{eq:NSM:E}).

%We note, that in applications involving intersticial atoms, we allow empty and occupied states of a site to define species of the site, cf.~\ref{sec:hydrogen_example}.

Enforcing the constraints (\ref{eq:NSM:E}) by means of Lagrange multipliers $k_B\{\beta\} \equiv (k_B \beta_i)_{i=1}^{N}$ and $k_B\{{\gamma}\}\equiv (k_B {\gamma}_i)_{i=1}^{N}$, with ${\gamma}_i = (\gamma_{ik})_{k=1}^M$, respectively, leads to the Lagrangian
\begin{equation}\label{eq:NSM:K}
\begin{split}
    &
    \mathcal{L}[\rho, \{\beta\}, \{{\gamma}\}]
    =
    \mathcal{S}[\rho]
    -
    k_B \{\beta\}^T \times \\ & \qquad\qquad
    \big(
        \{ \langle h \rangle \} - \{ e \})
        +
        k_B \{{\gamma}\}^T
        (\{\langle {n} \rangle \} - \{ \chi \}
    \big) ,
\end{split}
\end{equation}
where we write $\{ \langle h \rangle \} \equiv (\langle h_i \rangle)_{i=1}^{N}$, $\{ \langle {n} \rangle \} \equiv (\langle {n}_i \rangle)_{i=1}^N$, $\{ e \} \equiv (e_i)_{i=1}^{N}$, $\{ \chi \} \equiv (\chi_i)_{i=1}^N$. In view of identities (\ref{eq:NSM:N}), we additionally append the constraints
\begin{equation}
  \label{eq:NSM:chooseC}
  \sum_{k=1}^M
  \gamma_{ik}
  =
  0,
\end{equation}
in order to render $\{{\gamma}\}$ determinate.

Maximizing $\mathcal{L}[\cdot, \{\beta\}, \{{\gamma}\}]$ among probability measures gives
\begin{equation}\label{eq:NSM:GC}
    \rho(\{{q}\}, \{{p}\}, \{{n}\})
    =
    \frac{1}{{\Xi}}
    {\rm e}^{-\{\beta\}^T
    \{ h \}
    + \{ {\gamma}\}^T \{{n} \}
    } ,
\end{equation}
with
\begin{equation}\label{eq:NSM:Z}
    {\Xi}
    =
    \sum_{\{{n}\}\in\mathcal{O}_{N M}} \frac{1}{h^{3N}}
    \int_{\Gamma}
    {\rm e}^{
      -\{\beta\}^T \{ h \}
      +\{ {\gamma}\}^T \{{n} \}
    }
    \, dq \, dp .
\end{equation}
The corresponding equilibrium values of $\{\beta\}$ and $\{{\gamma}\}$ follow from (\ref{eq:NSM:E}) as a function of $\{e\}$ and $\{{x}\}$.

We interpret (\ref{eq:NSM:GC}) and (\ref{eq:NSM:Z}) as non-equilibrium generalizations of the {\sl Gibbs grand-canonical probability density function} and the {\sl grand-canonical partition function}, respectively,
\begin{equation}
    T_i = \frac{1}{k_B\beta_i} ,
    \qquad
    {\mu}_i
    =
    \frac{{\gamma}_i}{\beta_i}
    =
    k_B T_i {\gamma}_i,
\end{equation}
as the local {\sl absolute temperature} and {\sl chem\-ical-potential array} of particle $i$, respectively.

\subsection{Meanfield approximation} \label{Sec:MFA}

The calculation of the thermodynamic potentials in closed form is generally intractable and approximation is required. A variational meanfield theory \cite{Stanley-71, yeomans1992statistical} may be formulated by restricting (\ref{eq:NSM:K}) to some class of probability density functions of the form
\begin{equation}\label{eq:NSM:p0}
    \rho_0(\{{q}\}, \{{p}\}, \{{n}\})
    =
    \frac{1}{{\Xi}_0}
    {\rm e}^{-\{\beta\}^T
    \{ h_0 \}
    +
    \{ {\gamma} \}^T \{{n} \}
    } ,
\end{equation}
with
\begin{equation}\label{eq:NSM:Z0}
    {\Xi}_0
    =
    \sum_{\{{n}\}\in\mathcal{O}_{N M}} \frac{1}{h^{3N}}
    \int_{\Gamma}
    {\rm e}^{-\{\beta\}^T
    \{ h_0 \}
    +
    \{ {\gamma}\}^T \{{n} \}
    }
    \, dq \, dp ,
\end{equation}
and $\{ h_0 \}$ in some class $\mathcal{H}_0$ of {\sl local} trial Hamiltonians, possibly defined parametrically.

The restricted Lagrangian is
\begin{equation}\label{eq:NSM:K0}
\begin{split}
  &  \mathcal{L}_0[\rho_0, \{\beta\}, \{{\gamma}\}]
    =
  \mathcal{S}_0[\rho_0]
    - \\ &
    k_B \{\beta\}^T
    (\{ \langle h_0 \rangle_0 \} - \{ e \})
    +
    k_B \{{\gamma}\}^T
    (\{\langle {n} \rangle_0 \} - \{ \chi \}) ,
\end{split}
\end{equation}
where
\begin{equation}
    \mathcal{S}_0[\rho_0] = - k_B \langle \log \rho_0 \rangle_0
\end{equation}
and $\langle \cdot \rangle_0$ denotes averaging with respect to $\rho_0$. Inserting (\ref{eq:NSM:p0}) and (\ref{eq:NSM:Z0}) into (\ref{eq:NSM:K0}) using (\ref{eq:ESM:Sgeneral}) gives
\begin{equation}\label{eq:NSM:K0a}
\begin{split}
    &
    \mathcal{L}_0[\{h_0\}, \{\beta\}, \{\gamma\}]
    = \\ &
    k_B \{\beta\}^T \{ e \}
    -
    k_B \{{\gamma}\}^T \{ \chi \}
    -
    k_{\rm B} \log{\Xi}_0 .
\end{split}
\end{equation}

The optimal trial Hamiltonians are determined by maximizing $\mathcal{L}_0[\cdot, \{\beta\}, \{\gamma\}]$ with respect to $\{h_0\}$, or some suitable parametrization of $\{h_0\}$ thereof. In addition, the corresponding meanfield equilibrium values of $\{\beta\}$ and $\{{\gamma}\}$ follow as a function of $\{e\}$ and $\{{\chi}\}$ from the Euler-Lagrange equations of $\mathcal{L}_0[\{h_0\}, \{\beta\}, \{\gamma\}]$,
\begin{equation}\label{eq:mean_eq_val}
    \langle h_{0i} \rangle_0
    =
    e_i ,
    \qquad
    \langle n_{ik} \rangle_0
    =
    \chi_{ik} .
\end{equation}

\subsection{Interstitial hydrogen in metals}
\label{sec:hydrogen_example}

% cf. Sun, X.~and Ariza, M.~P.~and Ortiz, M.~and Wang, K.~G.~(2017) Acceleration of diffusive molecular dynamics simulations through mean field approximation and subcycling time integration. Journal of Computational Physics, 350, pp. 470-492.

As a special case of the general theory just outlined,
consider a crystal lattice where the base lattice sites are always occupied by metal atoms, while the interstitial sites are either occupied by an H atom, or unoccupied \cite{Sun:2017}. We index the base lattice sites by an index set $I_\text{M}$, and the interstitial sites by $I_\text{H}$. The occupancy number $n_i$ takes the fixed value of $1$, if $i\in I_\text{M}$ and the values of $0$ (site unoccupied) and $1$ (site occupied) if $ i\in I_\text{H}$. We assume a Hamiltonian of the form
\begin{equation}
\begin{split}
    & H(\{{q}\}, \{{p}\}, \{{n}\})
    = \\
    & \sum_{i\in{I_\text{M}}}
    \dfrac{1}{2m_\text{M}}|{p}_i|^2
    +
    \sum_{i\in{I_\text{H}}}
    \dfrac{1}{2m_\text{H}}|{p}_i|^2
    +
    V(\{{q}\}, \{{n}\}) ,
\end{split}
\end{equation}
where $V(\{{q}\}, \{{n}\})$ is an interatomic potential and $m_\text{M}$ and $m_\text{H}$ are the atomic masses of the metal host and hydrogen, respectively. We further assume that the system is in thermal equilibrium, whence $\beta_i = 1/k_{\rm B} T$.

We choose trial Hamiltonians for meanfield approximation of the form
\begin{equation}\label{eq:trialhami}
    \displaystyle \bar{h}_{i}({q}_i,{p}_i,n_i) =
    \qquad\qquad\qquad\qquad\qquad\qquad\quad
\end{equation}
\begin{equation} \nonumber
    \begin{cases}
        \dfrac{k_\text{B}T}{2\bar{\sigma}_i^2}|{q}_i - \bar{{q}}_i|^2
        +
        \dfrac{1}{2m_\text{M}}|{p}_i - \bar{{p}}_i|^2,
        & \text{if } i\in I_\text{M} , \\
        \dfrac{k_\text{B}T}{2\bar{\sigma}_i^2}|{q}_i - \bar{{q}}_i|^2
        +
        \dfrac{1}{2m_\text{H}}|{p}_i - \bar{{p}}_i|^2
        -
        k_\text{B}T \bar{\gamma}_{i}n_i,
        & \text{if } i\in I_\text{H} ,
    \end{cases}
\end{equation}
where $\bar{{q}}_i$, $\bar{\sigma}_i$, $\bar{{p}}_i$, and $\bar{\gamma}_{i}$, $i=1,2,\cdots,N$, are parameters that characterize the trial space,  $m_\text{M}$ and $m_\text{H}$ denote the atomic mass of the metal and hydrogen, respectively. The parameters $\bar{{q}}_i$ and $\bar{\sigma}_i$ are the mean and standard deviation of ${q}_i$, respectively, $\bar{{p}}_i$ is the mean value of ${p}_i$, and $\bar{\gamma}_{i}$ is the mean value of $\gamma_i$. From (\ref{eq:trialhami}), the corresponding meanfield probability density function (\ref{eq:NSM:p0}) evaluates to
\begin{equation}\label{eq:trialP_MH}
\begin{split}
&   \rho_0(\{{q}\}, \{{p}\}, \{{n}\})
    =
    \dfrac{1}{\mathnormal{\Xi}_0}
    \exp
    \Big\{ \\ &
    - \sum_{i\in{I_\text{M} \cup I_\text{H}}}\frac{1}{2\bar{\sigma}_i^2}|{q}_i - \bar{{q}}_i|^2 -
  \sum_{i\in{I_\text{M}}} \frac{\beta}{2 m_{\text{M}}}|{p}_i - \bar{{p}}_i|^2  \\ & -
        \sum_{i\in{I_\text{H}}}
        \Big( \frac{\beta}{2 m_{\text{H}}}|{p}_i - \bar{{p}}_i|^2
    -(\bar{\gamma}_{i}+\gamma_i) n_i \Big)
    \Big\},
\end{split}
\end{equation}
and the meanfield grand-canonical partition function (\ref{eq:NSM:Z0}) to
\begin{equation}\label{eq:trialpartfun_MH}
\begin{split}
\mathnormal{\Xi}_0
    =&
    \Big\{
        \prod_{i\in{I_\text{M}}}
        \Big(\dfrac{\bar{\sigma}_i\sqrt{ m_{\text{M}}/ \beta}}{\hbar}\Big)^3
    \Big\} \times \\ &
    \Big\{
        \prod_{i\in{I_\text{H}}}
        \Big(\dfrac{\bar{\sigma}_i\sqrt{ m_{\text{H}}/ \beta}}{\hbar}\Big)^3
        \Big( 1 + {\rm e}^{\bar{\gamma}_{i}+\gamma_i}\Big)
    \Big\} ,
\end{split}
\end{equation}
with $\hbar$ the reduced Planck constant.

Under the assumed isothermal conditions, the meanfield Lagrangian (\ref{eq:NSM:K0}) reduces to
\begin{equation}\label{eq:mean_Lag}
\begin{split}
    &
    \mathcal{L}_0[\{h_0\}, \{\gamma\}]
    = \\ &
    - k_{\rm B} \log{\Xi}_0
    +
    \frac{1}{T} \sum_{i\in{I_\text{M} \cup I_\text{H}}} e_i
    -
    k_B \{{\gamma}\}^T \{ x \} ,
\end{split}
\end{equation}
and the meanfield Euler-Lagrange equations (\ref{eq:mean_eq_val}) evaluate to
\begin{subequations}\label{eq:meanEL}
\begin{align}
 &   \langle \bar{h}_{i} \rangle_0
    =
    \langle V_i \rangle_0
    +
    \frac{1}{2m_i} |\bar{p}_i|^2
    +
    \frac{1}{2 \beta}
    =
    e_i ,\\
&    \langle n_i \rangle_0
    =
    \frac{{\rm e}^{\bar{\gamma}_{i}+\gamma_i}}{1 + {\rm e}^{\bar{\gamma}_{i}+\gamma_i}}
    =
    \chi_i ,
\end{align}
\end{subequations}
whereupon (\ref{eq:mean_Lag}) simplifies to
\begin{equation}
\begin{split}
&    \mathcal{L}_0[\{h_0\}, \{\gamma\}]
    =
   - k_{\rm B} \log{\Xi}_0
    + \\
&  k_B \beta \Big(
        \langle V \rangle_0
        +
        \sum_{i\in{I_\text{M} \cup I_\text{H}}}
        \frac{1}{2m_i} |\bar{p}_i|^2 \Big)
        + \frac{1}{2} N k_{\rm B}
    - \\
&    k_B \Big(
        \sum_{i \in I_\text{M}}
        \gamma_i
        +
        \sum_{i \in I_\text{H}}
        \gamma_i \frac{{\rm e}^{\bar{\gamma}_{i}+\gamma_i}}{1 + {\rm e}^{\bar{\gamma}_{i}+\gamma_i}}
    \Big),
\end{split}
\end{equation}
where $N = \#I_\text{M} + \#I_\text{H}$ is the total number of sites in the system. Using these identities, the meanfield optimality conditions evaluate to
\begin{subequations} \label{BQB5Ua}
\begin{align}
\label{eq:dL0_dq}
&\frac{\partial \mathcal{L}_0}{\partial \bar{q}_i}
    =
    \frac{\partial}{\partial \bar{q}_i}\langle V \rangle_0
    =
    \langle \frac{\partial V}{\partial q_i} \rangle_0
    =
    0 ,
    \\ &
    \frac{\partial \mathcal{L}_0}{\partial \bar{p}_i}
    =
    \frac{1}{m_i} \bar{p}_i
    =
    0 , \\
\begin{split}    \label{eq:dL0_ds}
&    \frac{\partial \mathcal{L}_0}{\partial \bar{\sigma}_i}
    =
    - \frac{3 k_{\rm B}}{\bar{\sigma}_i}
    +
    k_{\rm B} \beta \frac{\partial}{\partial \bar{\sigma}_i}\langle V \rangle_0
    = 0 ,
\end{split}
    \\ \label{eq:dL0_dg}
&    \frac{\partial \mathcal{L}_0}{\partial \bar{\gamma}_i}
    =
     k_{\rm B}
    \frac{{\rm e}^{\bar{\gamma}_i + \gamma_i}}{(1 + {\rm e}^{\bar{\gamma}_i + \gamma_i})^2}
    \Big( - 1 - {\rm e}^{\bar{\gamma}_i + \gamma_i} + \beta \frac{\partial \langle V \rangle_0}{\partial \chi_i} - \gamma_i \Big)
    =
    0 .
\end{align}
\end{subequations}
Combining (\ref{eq:dL0_dg}) and the second of (\ref{eq:meanEL}) we obtain the explicit identities
\begin{equation}\label{eq:gameq}
    \gamma_i = \beta \frac{\partial \langle V \rangle_0}{\partial \chi_i} - \frac{1}{1-\chi_i},
    \quad
    \bar{\gamma}_{i}
    =
    \log\Big(\frac{\chi_i}{1-\chi_i}\Big)
    -
    \gamma_i.
\end{equation}
We recall from (\ref{eq:ESM:av}) that
\begin{equation} \label{dY68tn}
\begin{split}
 &    \langle V \rangle_0 =
  \sum_{\{{n}\}\in\mathcal{O}_{NM}} \frac{1}{h^{3N}}
  \times\\ & \qquad
    \int_{\Gamma}
    V(\{{q}\}, \{{n}\}) \rho_0(\{{q}\}, \{{p}\}, \{{n}\})
    \, dq \, dp .
\end{split}
\end{equation}
In view of (\ref{eq:trialP_MH}) and (\ref{eq:gameq}), (\ref{dY68tn}) reduces to
\begin{equation} \label{PkrWe6}
\begin{split}
&   \langle V \rangle_0 (\{\bar{q}\},\{\bar{\sigma}\},\{\chi\})
    =
    \sum_{\{{n}\}\in\mathcal{O}_{NM}} \int V(\{{q}\}, \{{n}\}) \\
&   \Big\{
            \prod_{j\in{I_\text{M} \cup I_\text{H}}}
            \frac{1}{(\sqrt{2\pi} \ \bar{\sigma}_j)^3}
            \exp \Big(- \frac{1}{2\bar{\sigma}_j^2}|{q}_j - \bar{{q}}_j|^2 \Big)
    \Big\}  \\
&   \Big\{
            \prod_{j\in I_\text{H}}
             \Big(
                \frac{1}{\chi_j} - 1
            \Big)^{1-n_j}
            \, \chi_j
        \Big\} \, dq ,
\end{split}
\end{equation}
which fully defines the meanfield equations (\ref{eq:meanEL}).

We resort to two further approximations to evaluate (\ref{PkrWe6}) . In order to avoid occupancy sums of combinatorial complexity, we apply Jensen's inequality as an equality, with the result
\begin{equation}\label{eq:jensen_V}
\begin{split}
&    \langle V \rangle_0 (\{\bar{q}\},\{\bar{\sigma}\},\{\chi\})
    \approx
    \int
        V(\{{q}\}, \{{\chi}\})\\
& \Big\{
            \prod_{j\in{I_\text{M} \cup I_\text{H}}}
            \frac{1}{(\sqrt{2\pi} \ \bar{\sigma}_j)^3}
\exp \Big(
             -   \frac{1}{2\bar{\sigma}_j^2}|{q}_j - \bar{{q}}_j|^2
            \Big)
        \Big\}
    \, dq .
\end{split}
\end{equation}
In addition, we approximate the remaining integral over configuration space by means of numerical quadrature,
\begin{equation}\label{eq:V0Quad}
    \langle V \rangle_0(\{\bar{q}\},\{\bar{\sigma}\},\{\chi\})
    \approx
    \sum_{k=1}^{k_\text{max}} V(\{{q}\}_k, \{{\chi}\}) \, W_k ,
\end{equation}
where $\{{q}\}_k$ and $W_k$ are $3N$-dimensional quadrature points and weights, respectively. In view of the Gaussian weight in the integral (\ref{eq:jensen_V}), the optimal choice of quadrature rule is Hermitian quadrature, see~\ref{Hermitian-quadrature}. In addition, owing to the multiplicative form of the Gaussian weight in (\ref{eq:jensen_V}), the Hermitian quadrature points can be conveniently defined atom by atom. For every atom $i$, the resulting Hermitian quadrature points are centered at $\bar{q}_i$ with relative positions rescaled by $\bar{\sigma}_i$.

Eq.~\eqref{eq:V0Quad} supplies an explicit expression for the meanfield interatomic potential. The corresponding meanfield equations follow from (\ref{BQB5Ua}) as
\begin{subequations}
\begin{align}
    0
    & =
    \sum_{k=1}^{k_\text{max}} \frac{\partial V(\{{q}\}_k, \{{\chi}\})}{\partial q_i} \, W_k ,
    \\
    0 & = \frac{1}{\bar{\sigma}_i^3}
    \sum_{k=1}^{k_\text{max}} \frac{\partial V(\{{q}\}_k, \{{\chi}\})}{\partial \{{q}\}_k} \cdot \frac{\partial \{{q}\}}{\bar{\sigma}_i} \, W_k - \frac{3\ k_{\rm B}}{\bar{\sigma}_i},
    \\
\gamma_i &= \beta \sum_{k=1}^{k_\text{max}} \frac{\partial V}{\partial\chi_i}(\{{q}\}_k, \{{\chi}\}) \, W_k  - \frac{1}{1-\chi_i},
\end{align}
\end{subequations}
which define a system of nonlinear equations that determine the value of $\{\bar{q}\}$, $\{\bar{\sigma}\}$ and $\{\chi\}$ at equilibrium.

\subsection{Higher-order meanfield approximation}

The simple meanfield approximation put forth in the foregoing can be systematically improved upon by recourse to cluster expansions \cite{Stanley-71, Binney-92}. In the present framework, such expansions may be conveniently formulated in terms of Hermite polynomials and multipole approximation formulas. A succinct summary of the theory is included next for completeness.

We begin by rewriting (\ref{eq:trialP_MH}) as
\begin{equation}
\begin{split}
&    \rho_0(\{{z}\}, \{{n}\})
    = \\
&    \dfrac{1}{\mathnormal{\Xi}_0}
    \exp
    \Big\{
        - \frac{1}{2} \{z'\}^T Q \{z'\}
        +
        \sum_{i\in{I_\text{H}}}
            (\bar{\gamma}_{i}+\gamma_i) n_i
    \Big\},
\end{split}
\end{equation}
with $z_i \equiv (q_i,p_i)$, $\bar{z}_i = (\bar{q}_i,\bar{p}_i)$, $z_i' = z_i - \bar{z}_i$ and
\begin{equation}
\begin{split}
    &
    \{z'\}^T Q \{z'\}
    \equiv \\ &
    \sum_{i=1}^N \frac{1}{2\bar{\sigma}_i^2}|{q}_i - \bar{{q}}_i|^2
    +
    \sum_{i=1}^N \frac{1}{2k_{\text{B}}T m_i}|{p}_i - \bar{{p}}_i|^2 .
\end{split}
\end{equation}
For $m = 0, 1, 2, \dots$, we may then generalize (\ref{eq:trialP_MH}) by means of the sequence of meanfield probability density functions
\begin{equation}\label{x2IyOF}
    \rho_m(\{{z}\}, \{{n}\})
    =
    P_m(\{{z'}\}) \rho_0(\{{z}\}, \{{n}\}) ,
\end{equation}
where
%\begin{equation}\label{2h0kP4}
%\begin{split}
%    &
%    P_m(\{{z}\})
%    =
%    \sum_{|\alpha|\leq m}
%    c_\alpha
%    H_\alpha(\{{z}\})
%    = \\ &
%    \sum_{|\alpha|\leq m}
%    c_\alpha
%    \exp\left( \frac{1}{2} \{{z}\}^T Q \{{z}\} \right)
%   D^\alpha
%    \exp\left( - \frac{1}{2} \{{z}\}^T Q \{{z}\} \right)
%\end{split}
%\end{equation}
\begin{equation}\label{2h0kP4}
    P_m(\{{z}\})
    =
    \sum_{|\alpha|\leq m}
    c_\alpha
    H_\alpha(\{{z}\}) ,
\end{equation}
is a linear combination of Hermite polynomials
\begin{equation}
\begin{split}
    &
    H_\alpha(\{{z}\})
    = \\ &
    \exp\Big( \frac{1}{2} \{{z}\}^T Q \{{z}\} \Big)
    D^\alpha
    \exp\Big( - \frac{1}{2} \{{z}\}^T Q \{{z}\} \Big)
\end{split}
\end{equation}
of degree less or equal to $m$. It is readily verified that $\rho_m(\{{z}\}, \{{n}\})$ sums to $1$ and, therefore, defines a probability density, provided that $c_0 = 1$.

We recall that Hermite polynomials are orthogonal polynomials with respect to Gaussian weights and, hence, parameterized distributions of the type (\ref{x2IyOF}) are dense in a suitably weighted $L^2$ space \cite{Courant:1989} and can be used to approximate the true probability density function $\rho(\{{z}\}, \{{n}\})$ to any degree of accuracy.

The remainder of the meanfield model follows {\sl mutatis mutandis} Section~\ref{sec:hydrogen_example}, with (\ref{dY68tn}) replaced by
\begin{equation}
\begin{split}
    &
    \langle V \rangle_0
    =
    \sum_{\{{n}\}\in\mathcal{O}_{NM}} \frac{1}{h^{3N}}
    \times \\ & \qquad
    \int_{\Gamma}
    V(\{{q}\}, \{{n}\})
    \;
    \rho_m(\{{z}\}, \{{n}\} \})
    \, dz .
\end{split}
\end{equation}
As before, because of the Gaussian weights and the orthogonality properties of the Hermite polynomials, the requisite phase-space integrals can be conveniently approximated by recourse to Hermitian quadrature, \ref{Hermitian-quadrature}.

\section{Angular-Dependent Potential}
\label{sec:potential}

In the remainder of the paper, we specialize the preceding theoretical framework  to Angular-Dependent interatomic Potentials (ADP), including meanfield approximation. ADP is a class of many-body potentials developed by \cite{MISHIN2006} as an extension of the Embedded-Atom Method (EAM) potentials \cite{daw1984embedded}. For binary systems such as metal-hydrogen (M-H), assuming the occupancy of the metal sites to be fixed at $n_i = 1$ and the occupancy of the hydrogen sites to vary over the full range $n_i \in \left(0, 1\right)$, the ADP potential takes the form
\begin{equation}\label{eq:adp}
\begin{split}
&    V(\{{q}\} , \{n\}) =
{V^{F}(\{{q}\} , \{n\})} + {V^{\phi}(\{{q}\} , \{n\})} + \\ &
{V^{\mu}(\{{q}\} , \{n\})} +{V^{\lambda}(\{{q}\} , \{n\})} - {V^{tr\lambda}(\{{q}\} , \{n\})} , 
\end{split}
\end{equation}
where
\begin{equation}\label{eq:adpF}
V^{F}(\{{q}\} , \{n\}) =
\sum_{i \in I_\text{M}} F_\text{M} (\bar{\rho}_{i}) + \sum_{i \in I_\text{H}} n_i F_\text{H} (\bar{\rho}_{i}) ,
\end{equation}
is the embedding energy,
\begin{equation}\label{eq:adpPhi}
\begin{split}
& V^{\phi}(\{{q}\} , \{n\}) = \\ &
\frac{1}{2}\sum_{\substack{i , j \in I_\text{M} \\ j \neq i}} \phi_\text{MM} +
\frac{1}{2}  \sum_{\substack{i \in I_\text{H}, j \in I_\text{M} \\ j \neq i}} n_i \phi_\text{HM} + \\ &
\frac{1}{2} \sum_{\substack{i \in I_\text{M}, j \in I_\text{H} \\ j \neq i}} n_j \phi_\text{MH} +
\frac{1}{2} \sum_{\substack{i, j \in I_\text{H} \\ j \neq i}} n_i n_j \phi_\text{HH}
\end{split}
\end{equation}
is the pairwise-interaction energy,
\begin{equation}\label{eq:adpMu}
\begin{split}
    &
    V^{\mu}(\{{q}\} , \{n\}) = \\ &
    \frac{1}{2} \sum_{i \in I_\text{M}} \left(\mu^\text{M}_i \cdot \mu^\text{M}_i \right)\frac{1}{2} +
    \sum_{i \in I_\text{H}} n_i^2\ \left(\mu^\text{H}_i \cdot \mu^\text{H}_i \right)
\end{split}
\end{equation}
is the energy due to dipole distortion, and
\begin{equation}\label{eq:adpLambda}
\begin{split}
&
V^{\lambda}(\{{q}\} , \{n\}) = \\ &
\frac{1}{2} \sum_{i \in I_\text{M}} \left(\lambda^\text{M}_i \colon \lambda^\text{M}_i \right) +
\frac{1}{2} \sum_{i \in I_\text{H}} n_i^2\ \left(\lambda^\text{H}_i \colon \lambda^\text{H}_i\right)
\end{split}
\end{equation}
and
\begin{equation}\label{eq:adpTr}
\begin{split}
&
V^{\text{tr}\lambda}(\{{q}\} , \{n\}) = \\ &
- \frac{1}{6} \sum_{i \in I_\text{M}} \left(\text{tr} \lambda^\text{M}_i\right)^2 - \frac{1}{6} \sum_{i \in I_\text{H}}  n_i^2 \left(\text{tr} \lambda^\text{H}_i\right)^2
\end{split}
\end{equation}
are energies accounting for quadrupole distortions.

In (\ref{eq:adpF}), $F_\text{M}\ (\bar{\rho}_{i})$ and $F_\text{H}\ (\bar{\rho}_{i})$ are the embedding energy functions for metal and hydrogen atoms, respectively. The embedding energies are a function of the electron density
\begin{equation}
\bar{\rho}_i =
\sum_{j\in{I_\text{M}}, ~j\neq i}\rho_\text{M}(r_{ij})
+ \sum_{j\in{I_\text{H}}, ~j\neq i}n_j \rho_\text{H}(r_{ij}),
\label{eq:sumelectron}
\end{equation}
with $r_{ij}=|{q}_i - {q}_j|$, which combines contributions $\rho_\text{M}(r_{ij})$ and $\rho_\text{H}(r_{ij})$ from metal and hydrogen atoms, respectively. In (\ref{eq:adpPhi}), $\phi_\text{MM}$, $\phi_\text{MH}$, and $\phi_\text{HH}$ are pair potentials between metal-metal, metal-hydrogen, and hydrogen-hydrogen atom pairs, respectively. The non-central interactions, (\ref{eq:adpMu}), (\ref{eq:adpLambda}) and (\ref{eq:adpTr}), account for the effect of dipole distortions
\begin{subequations}
\begin{align}
&\mu^\text{M}_{i} = \sum_{\substack{j \in I_{\text{M}} \\ j \neq i}} u_{\text{MM}} (r_{ij}) {r}_{ij} +\sum_{\substack{j \in I_{\text{H}} \\ j \neq i}} n_j u_{\text{MH}} (r_{ij}) {r}_{ij},\\
&\mu^\text{H}_{i} = \sum_{\substack{j \in I_{\text{M}} \\ j \neq i}} u_{\text{HM}} (r_{ij}) {r}_{ij}
+ \sum_{\substack{j \in I_{\text{H}} \\ j \neq i}} n_j u_{\text{HH}} (r_{ij}) {r}_{ij}.
\end{align}
\end{subequations}
and quadrupole distortions
\begin{subequations}\label{eq:quadru-distortion-species}
\begin{align}
\lambda^\text{M}_i &= \sum_{\substack{j \in I_{M} \\ j \neq i}} w_{\text{MM}} (r_{ij}) \left({r}_{ij} \otimes {r}_{ij}\right) \\ \nonumber
& +\sum_{\substack{j \in I_{H} \\ j \neq i}} n_j w_{\text{MH}}(r_{ij}) \left({r}_{ij} \otimes {r}_{ij}\right),\\
\lambda^\text{H}_i &= \sum_{\substack{j \in I_{M} \\ j \neq i}} w_{\text{HM}} (r_{ij}) \left({r}_{ij} \otimes {r}_{ij}\right)  \\ \nonumber
&+\sum_{\substack{j \in I_{H} \\ j \neq i}} n_j w_{\text{HH}}(r_{ij}) \left({r}_{ij} \otimes {r}_{ij}\right).
\end{align}
\end{subequations}
The requisite functions $F_M$, $F_H$, $\phi_{\text{MM}}$, $\phi_{\text{HM}}$, $\phi_{\text{HH}}$, $\phi_{\text{MM}}$, $w_{\text{MH}}$ and $w_{\text{HH}}$$u_{\text{MM}}$, $u_{\text{HM}}$, $u_{\text{HH}}$ and $w_{\text{MM}}$, $w_{\text{MH}}$ and $w_{\text{HH}}$ are commonly represented in tabular form, e.~g.~\cite{SMIRNOVA}.

\subsection{Meanfield approximation of the ADP potential}
\label{sec:ADPmeanfield}

The thermalized ADP potential $\meanfield{V(\{{q}\}, \{\chi\})}$ follows by inserting (\ref{eq:adp}) into (\ref{eq:jensen_V}). We streamline computations by means of approximations of the form
\begin{equation}
\label{eq:mf-embedding-term-approximation}
    \langle F_\text{Mg}(\bar{\rho}_i)\rangle_0
    \approx
    F_\text{Mg}( \langle\bar{\rho}_i\rangle_0) ,
\end{equation}
which correspond to saturating Jensen's inequality for convex functions as an equality.

In addition, the direct evaluation of the angular terms proves unwieldy and requires further streamlining. To this end, we reduce the angular terms to a sum over triads by considering the permutations $i$--$j_1$--$j_2$. A straightforward derivation gives
\begin{equation} \label{eq:dipole-distortion-term-II}
\begin{split}
&\meanfield{V^{{\mu}}} = \frac{1}{2} \sum_{i \in I_\text{Mg} \cup I_\text{H} } \sum_{\substack{j_1,j_2 \in \mathcal{B}_i(r_\text{c}) \\ j_1,j_2 \neq i}}  \meanfield{{\mu}_{ij_1} \cdot {\mu}_{ij_2}},
\end{split}
\end{equation}
for the dipole contribution,
\begin{equation}\label{eq:quadrupole-distortion-term-II}
\begin{split}
&\meanfield{V^{{\lambda}}} =\frac{1}{2} \sum_{i \in I_\text{Mg} \cup I_\text{H} } \sum_{\substack{j_1,j_2 \in \mathcal{B}_i(r_\text{c}) \\ j_1,j_2 \neq i}} \meanfield{{\lambda}_{ij_1} : {\lambda}_{ij_2}},
\end{split}
\end{equation}
for the quadrupole contribution,
\begin{equation}\label{eq:tr-quadrupole-distortion-term-II}
\begin{aligned}
&\meanfield{V^{\text{tr}{\lambda}}} = \frac{1}{6}\ \sum_{i \in I_\text{Mg} \cup I_\text{H} } \sum_{\substack{j_1,j_2 \in \mathcal{B}_i(r_\text{c}) \\ j_1,j_2 \neq i}} \meanfield{\text{tr}{\lambda}_{ij_1}\, \text{tr}{\lambda}_{ij_2}}.
\end{aligned}
\end{equation}
and for trace of the quadrupole contribution. In these expressions the combination of species have been omitted for the sake of clarity.

\subsection{Numerical quadrature}

Proceeding as in (\ref{eq:V0Quad}), we evaluate meanfield expectations $\langle \cdot \rangle_0$ by recourse to numerical quadrature. For instance, an application of Hermitian quadrature gives
\begin{equation}\label{eq:embedgaussnew1}
\begin{split}
\langle \bar{\rho}_i \rangle_0
& \approx
\sum_{\substack{j\in{I_\text{Mg}}\cap \mathcal{B}_i(r_\text{c}) \\ j\neq i}}
\sum_{k=1}^{k_{max}} \rho_\text{Mg}(r_{ij,k}) W_k
\\ & +
\sum_{\substack{j\in{I_\text{H}}\cap \mathcal{B}_i(r_\text{c}) \\ j\neq i}} \chi_j
 \sum_{k=1}^{NQ} \rho_\text{H}(r_{ij,k}) W_k
\end{split}
\end{equation}
and likewise for the pair-potential terms. In these expressions, $\mathcal{B}_i(r_\text{c})$ denotes the set of atomic sites within a distance $r_c$ of site $i$, and $r_\text{c}$ is the cut-off distance of the ADP potential. The subscript $k$ on $r$ refers to the $k$-th Gaussian point, with $W_k$ the corresponding weight, and $k_\text{max}$ denotes the total number of Gaussian points on a pre-defined sparse grid, which is at least $2n$ for $n$-dimensional integration. For Eq.~(\ref{eq:embedgaussnew1}), we have
\begin{equation}
\begin{split}
    &
    \Big|\big(I_{\text{Mg}}\cap \mathcal{B}_i(r_\text{c})\big)\cup \big(I_{\text{H}}\cap \mathcal{B}_i(r_\text{c})\big)\Big|
    = \\ &
    \Big|\big(I_{\text{Mg}}\cup I_{\text{H}}\big) \cap \mathcal{B}_i(r_\text{c})\Big| = Q+1,
\end{split}
\label{eq:defq}
\end{equation}
and
\begin{equation}
k_\text{max} \geq 6(Q+1),
\end{equation}
since the Gaussian quadratures are performed with respect to 6-dimensional integration. Here, $k_\text{max}$ is independent of $Q$ and $N$, where $N$ is the total number of base and interstitial sites in the material sample.

Therefore, the evaluation of the pairwise terms in (\ref{eq:adp}) requires $\mathcal{O}(QN)$ calls of pair functions.

By contrast, a similar analysis shows that the numerical integration by Hermitian quadrature of the angular terms in (\ref{eq:adp}) requires $\mathcal{O}(Q^2N)$ calls. In addition, we find that the accurate integration of the angular terms requires the use of Hermitian quadrature of higher order, which adds further to the computational expense. Under these conditions, the method of multipole expansion, \ref{Multipole-quadrature}, a general scheme for the approximation of integrals that is well-suited to the evaluation of phase-space averages, suggests itself as an alternative.

\subsection{Computational efficiency}

\begin{figure*}[htb]
\begin{center}
    \subfigure[]{\includegraphics[width=8cm]{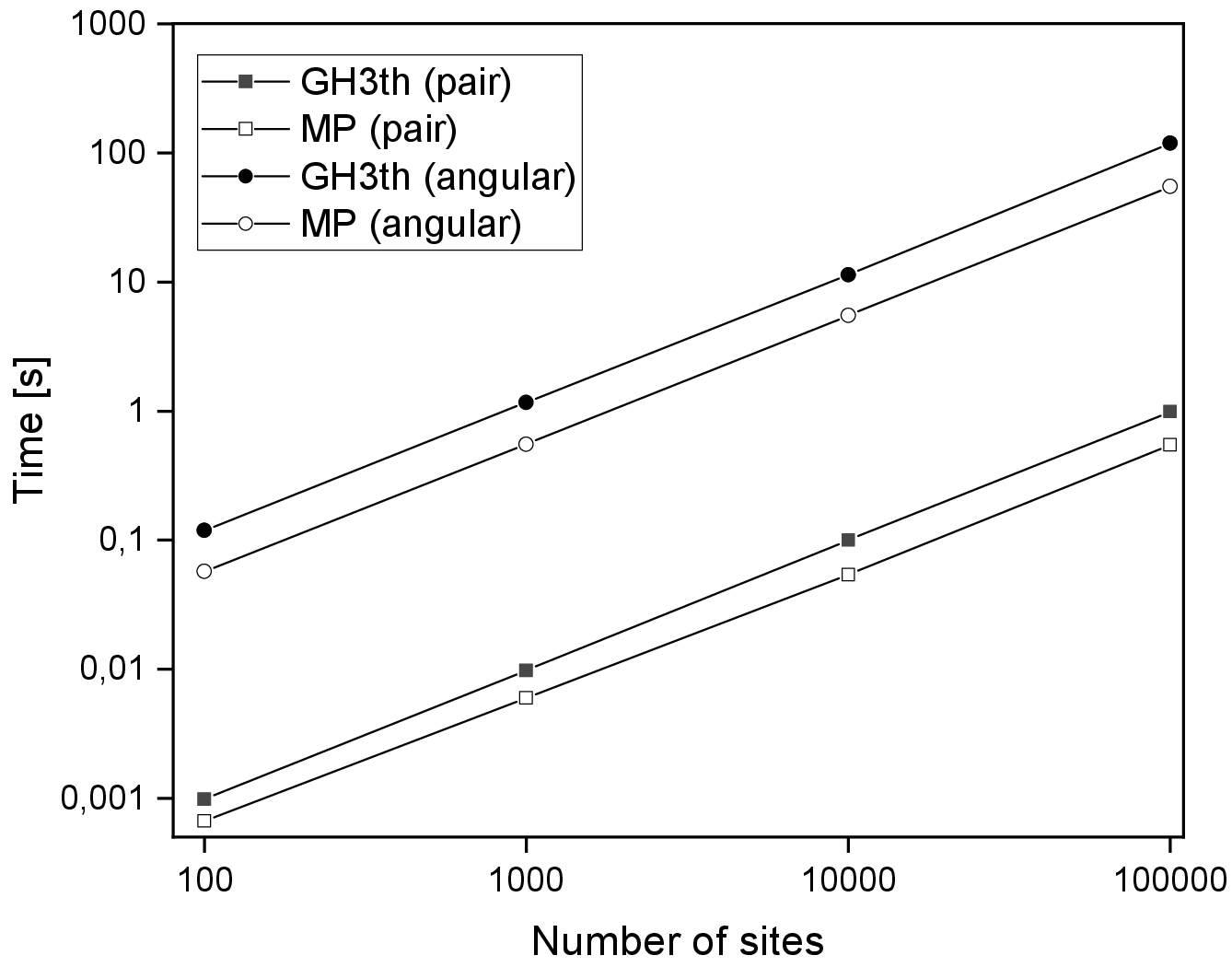}}
    \subfigure[]{\includegraphics[width=7.8cm]{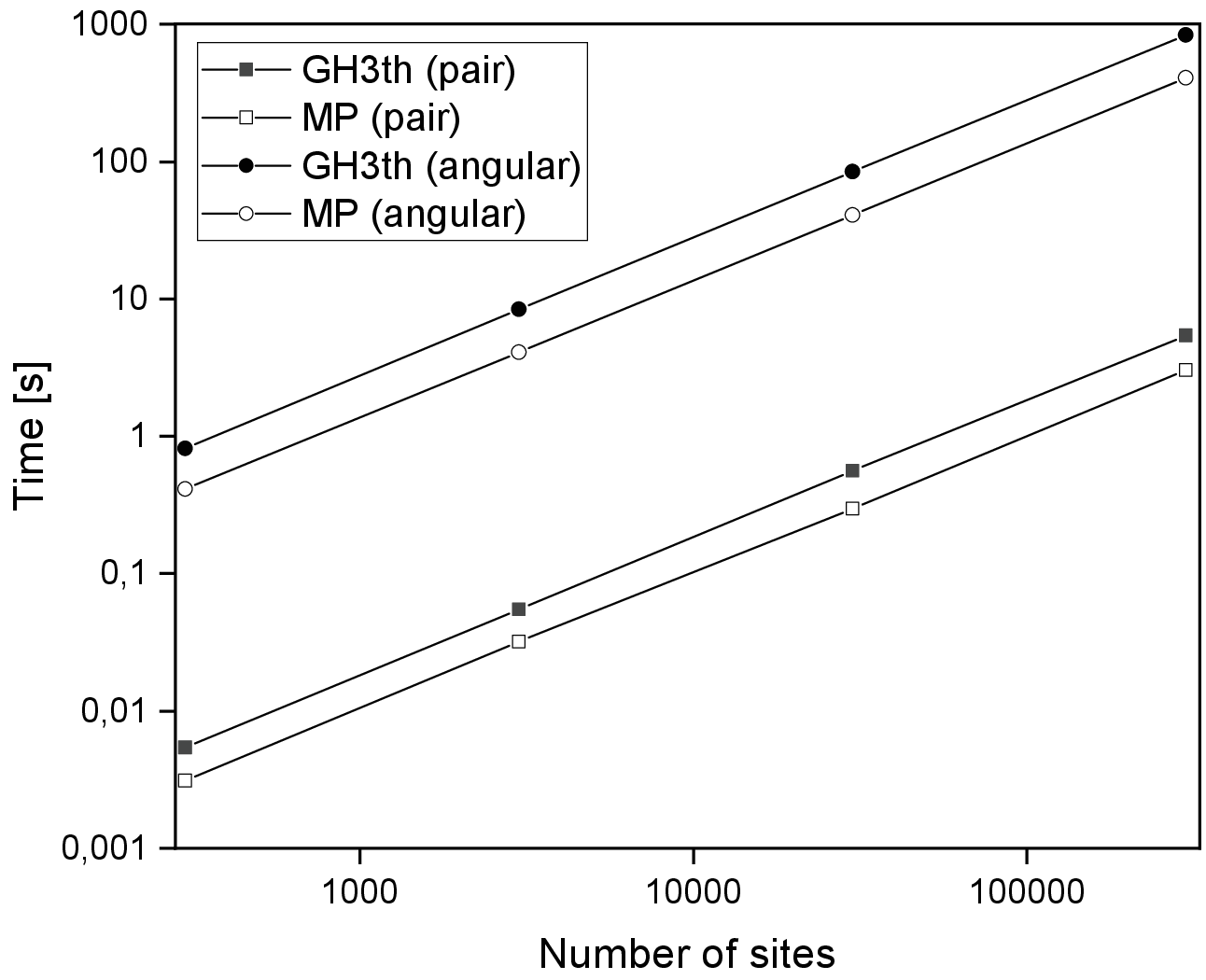}}
    \caption{Elapsed computation time. (a) Magnesium, hexagonal lattice structure. (b) Magnesium hydride MgH$_2$, rutile lattice structure.} \label{figtime}
\end{center}
\end{figure*}

In order to verify the implementation of the meanfield approximation of the ADP potential for magnesium hydrides and assess its numerical performance we have monitored the computational time required to evaluate $\meanfield{V(\{{q}\}, \{\chi\})}$ over a range of computational cells comprising 100 to 300000 sites. Calculations are performed at room temperature and both magnesium with hexagonal lattice structure and its hydride with rutile structure are evaluated.
Moreover, we record separately calculational times of pairwise and angular interactions. The elapsed time, measured in seconds, is plotted against the number of sites in the simulation in Fig.~\ref{figtime}. These plots reveal that second-order multipole quadrature (MP) is slightly superior to third order Gauss-Hermite (GH3th) quadrature in terms of computational efficiency.

\section{Validation of the meanfield approximation}\label{results}

We proceed to assess the ability of the thermalized ADP (TADP) potential to reproduce the equilibrium properties of magnesium (Mg) and one of its hydrides (MgH$_2$), including free entropy, heat capacity, thermal expansion, molar volumes, volumetric equation of state and elastic constants. In order to simulate bulk conditions, we consider two cubic domains subject to periodic boundary conditions, one for HCP Mg and another for rutile MgH$_2$. The unit lattice cells are looked up from the COD database, see~\cite{COD_data_base}. Each domain contains $3 \times 3 \times 3$ unit lattice cells encompassing a computational cell larger than the cut-off radius of the potential~\cite{SMIRNOVA}, as required. In all calculations, we employ second-order multipole quadrature (MP) and third-order Gauss-Hermite quadrature (GH3th) for the evaluation of meanfield expectations in order to assess sensitivity of the model to numerical integration.

\subsection{Thermodynamic properties}

\begin{figure}[htb]
    \begin{center}
    \includegraphics[width=0.6\columnwidth]{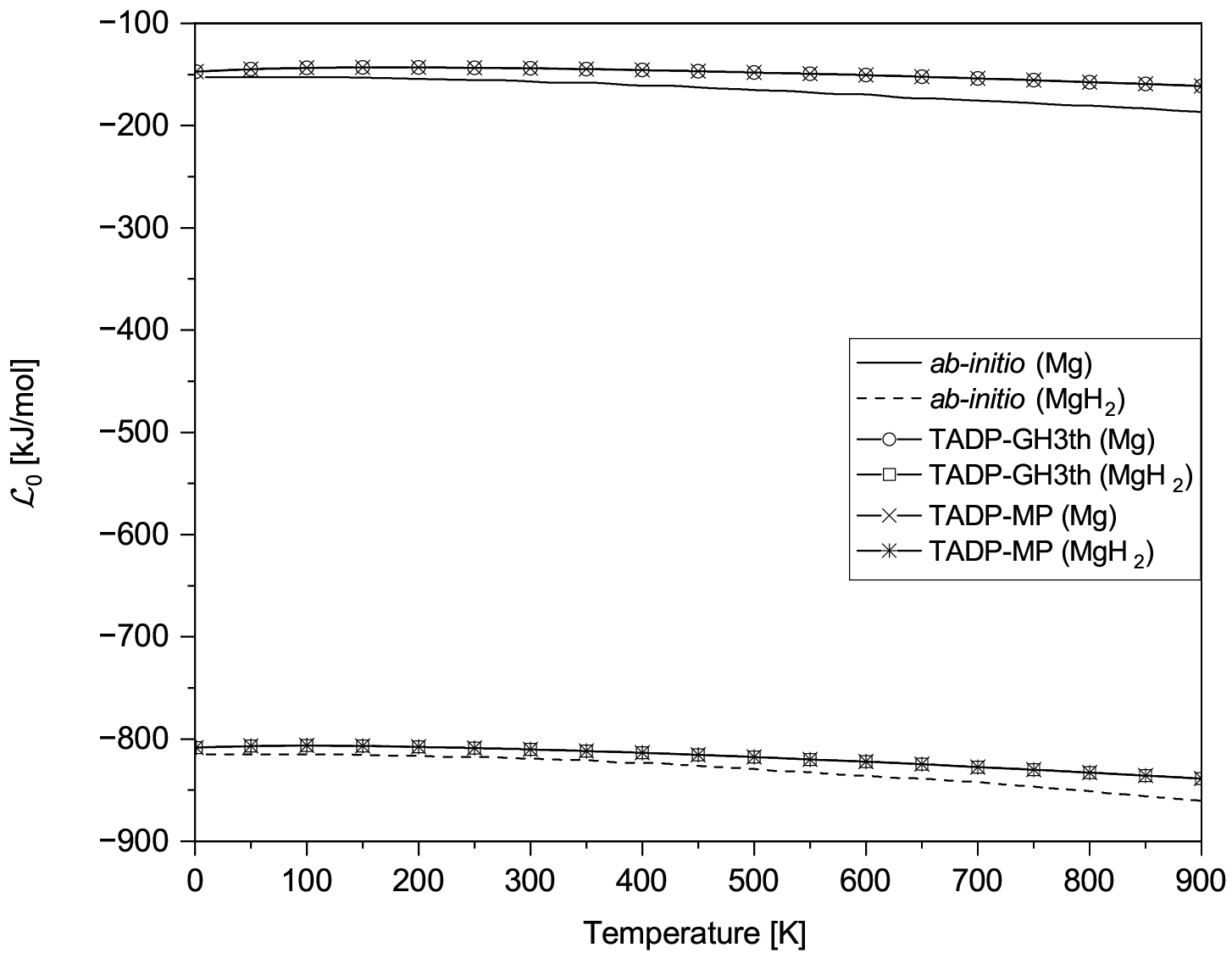}
    \end{center}
       \caption{HCP Mg and rutile MgH$_2$. Calculated dependence of the free entropy on temperature at zero pressure using second-order multipole quadrature (MP) and third-order Gauss-Hermite quadrature (GH3th). {\sl Ab initio} calculations of \cite{Junkaew_et_al_2014} also shown for comparison.}
    \label{figfreeentropy}
\end{figure}

We consider material samples undergoing affine deformations $F$ at temperature $T$, see \cite{Marsden:1994, Weiner:2002} for background and notation. We further consider an NPT ensemble characterized by the constrained free entropy
\begin{equation}
    \label{eq:thermal-expansion}
       \arg \max_{{F}, \bar{\sigma}}
       \{
           T \mathcal{L}_0 (\text{T}, \{h_0\}, \{\gamma\})
           -
           V_0 p J
       \} ,
\end{equation}
where $V_0$ is the undeformed volume of a reference unit cell, $p$ is a macroscopic pressure and $J = \det(F)$ is the Jacobian of the deformation. The corresponding Euler-Lagrange equations are
\begin{align}
    &\frac{\text{T}}{\text{V}_0}\ \frac{\partial \mathcal{L}_0(\bar{{q}},\bar{\sigma})}{\partial {F}} = p J \, F^{-T}, \\ &
    \frac{\text{T}}{\text{V}_0}\ \frac{\partial \mathcal{L}_0(\bar{{q}},\bar{\sigma})}{\partial \bar{\sigma}_i}\, =\, 0 ,
\end{align}
which are solved using a critical point line search \cite{Brune2015} implemented in the PETSc/TAO library~\cite{petsc-web-page} with $(pJ,T)$ as control parameters.

Fig.~\ref{figfreeentropy} concerns the dependence of the free entropy on temperature at zero pressure. As can be seen from the figure, the TADP values of the free entropy and the general decreasing trend of the free entropy with temperature are in good agreement with the {\it ab initio} calculations of \cite{Junkaew_et_al_2014} over a broad range of temperatures.

% \subsection{Heat capacity}

\begin{figure}
\centering
    \includegraphics[width=0.6\columnwidth]{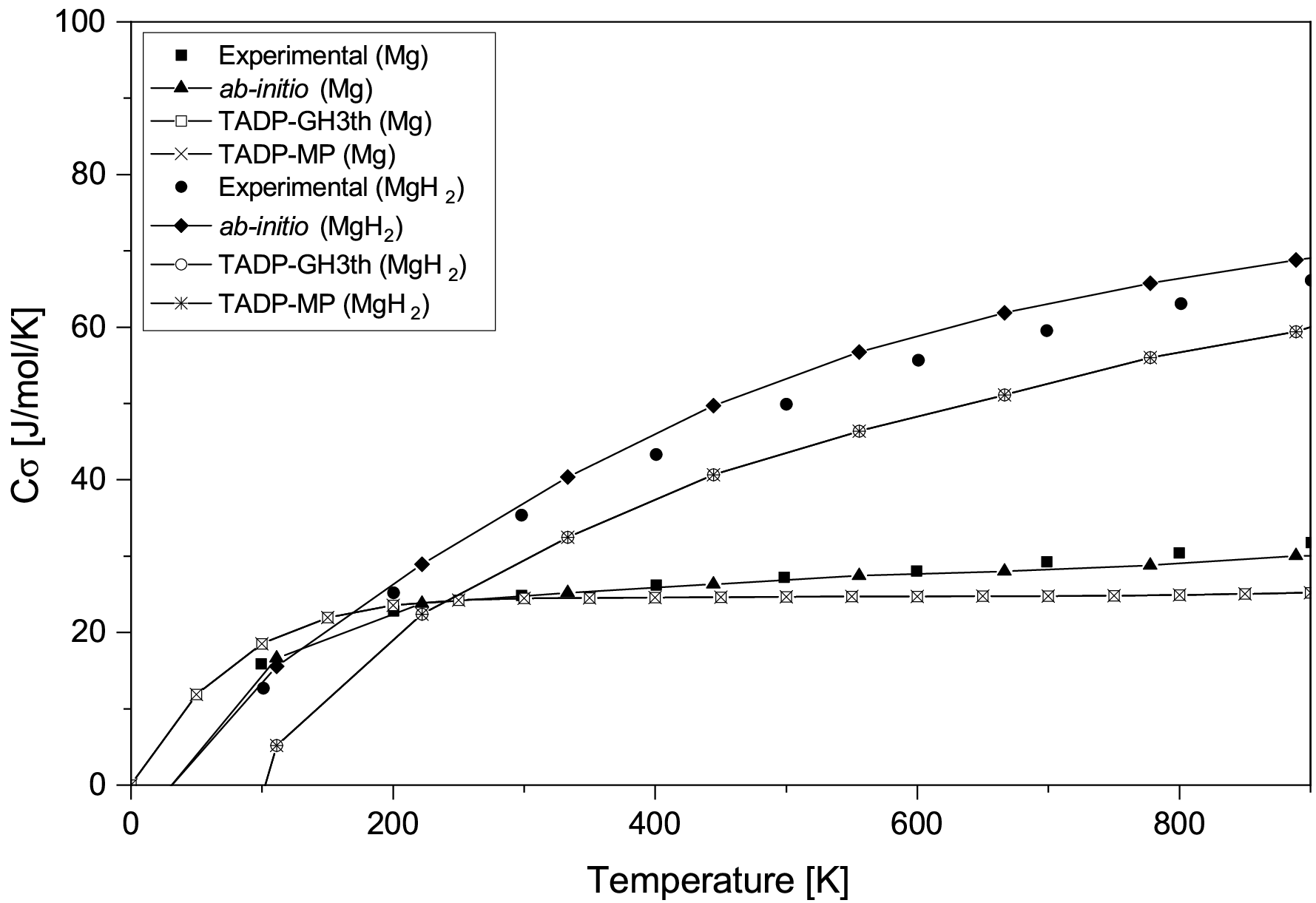}
    \caption{HCP Mg and rutile MgH${}_2$. Calculated dependence of the
    specific heat capacity on temperature at zero pressure
    using second-order multipole quadrature (MP) and third-order Gauss-Hermite quadrature (GH3th). Ab initio calculations of \cite{Junkaew_et_al_2014} and experimental data from \cite{Chase1998} also shown for comparison.}
\label{figspecificheat}
\end{figure}

Fig.~\ref{figspecificheat} shows the computed values of heat capacity at zero pressure over a range of temperatures, together with corresponding {\it ab initio} calculations \cite{Junkaew_et_al_2014} and experimental measurements \cite{Chase1998} by way of comparison. The good overall fidelity of the TADP predictions is clearly evinced by the comparison.

\subsection{Thermal expansion coefficient}

\begin{figure}
    \centering
    \includegraphics[width=0.6\columnwidth]{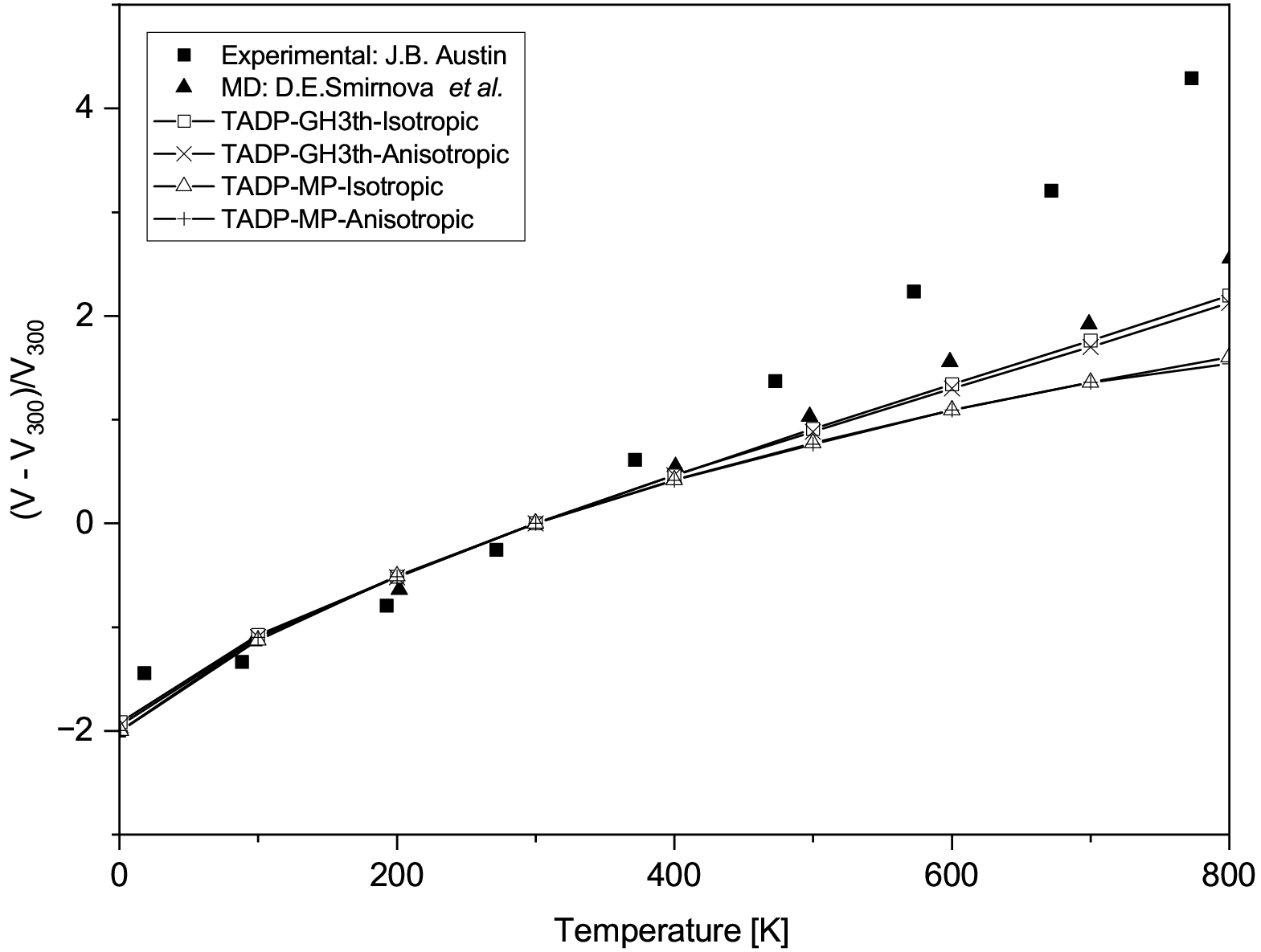}
    \caption{HCP Mg. Calculated dependence of the thermal expansion coefficient on temperature at zero pressure using second-order multipole quadrature (MP) and third-order Gauss-Hermite quadrature (GH3th). Molecular dynamics calculations of \cite{SMIRNOVA} and experimental data of \cite{Austin1932} also shown for comparison.}
    \label{figthermalexpansion}
\end{figure}

Fig.~\ref{figthermalexpansion} compares the computed dependence of the thermal expansion coefficient of HCP Mg {at zero pressure} on temperature, with and without isotropic approximation. {The predictions of TADP are verified against} MD calculations reported by \cite{SMIRNOVA} and {validated against} experimental data {from} \cite{Austin1932}. As may be seen from the figure, TADP agrees closely with MD over the entire temperature range under consideration, which tends to verify the statistical mechanical paradigm. In addition, TADP and MD match closely the experimental data up to roughly 400K, but increasingly deviate thenceforth. This discrepancy is suggestive of accuracy limitations of ADP potentials of \cite{SMIRNOVA} at high temperatures.

\begin{figure}[htb]
    \centering
    \includegraphics[width=0.6\columnwidth]{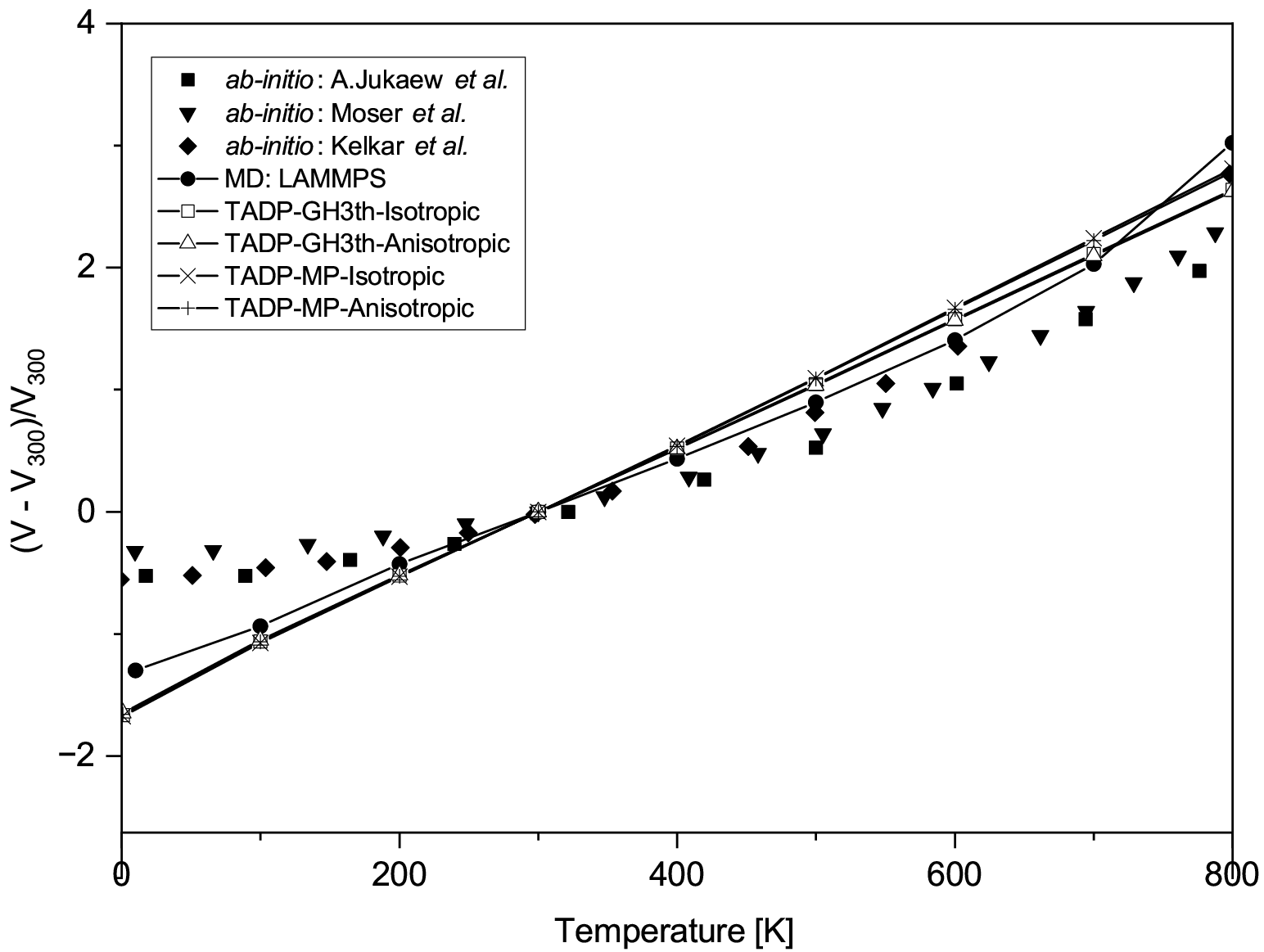}
    \caption{Rutile MgH$_2$. Calculated dependence of the thermal expansion coefficient on temperature at zero pressure using second-order multipole quadrature (MP) and third-order Gauss-Hermite quadrature (GH3th). Molecular dynamics calculations of \cite{SMIRNOVA} and {\it ab initio} calculations \cite{Junkaew_et_al_2014,  Kelkar_et_al_2008,  Moser_et_al_2011} also shown for comparison.}
    \label{figthermalexpansionrut}
\end{figure}

Fig.~\ref{figthermalexpansionrut} shows a similar comparison of the computed temperature dependence of the thermal expansion coefficient of rutile MgH$_2$ {at zero pressure} with MD and {\sl ab initio} calculations \cite{Junkaew_et_al_2014, Moser_et_al_2011, Kelkar_et_al_2008}. As may be seen from the figure, TADP closely matches MD over the entire temperature range, and {\sl ab initio} data above $\sim 300$K, but, as expected, deviates from {\it ab initio} at low temperature where quantum mechanical effects are non-negligible.

%%%%%%%%%%

\subsection{Isothermal compression curve}

\begin{figure}[htb]
    \centering
    \includegraphics[width=0.6\columnwidth]{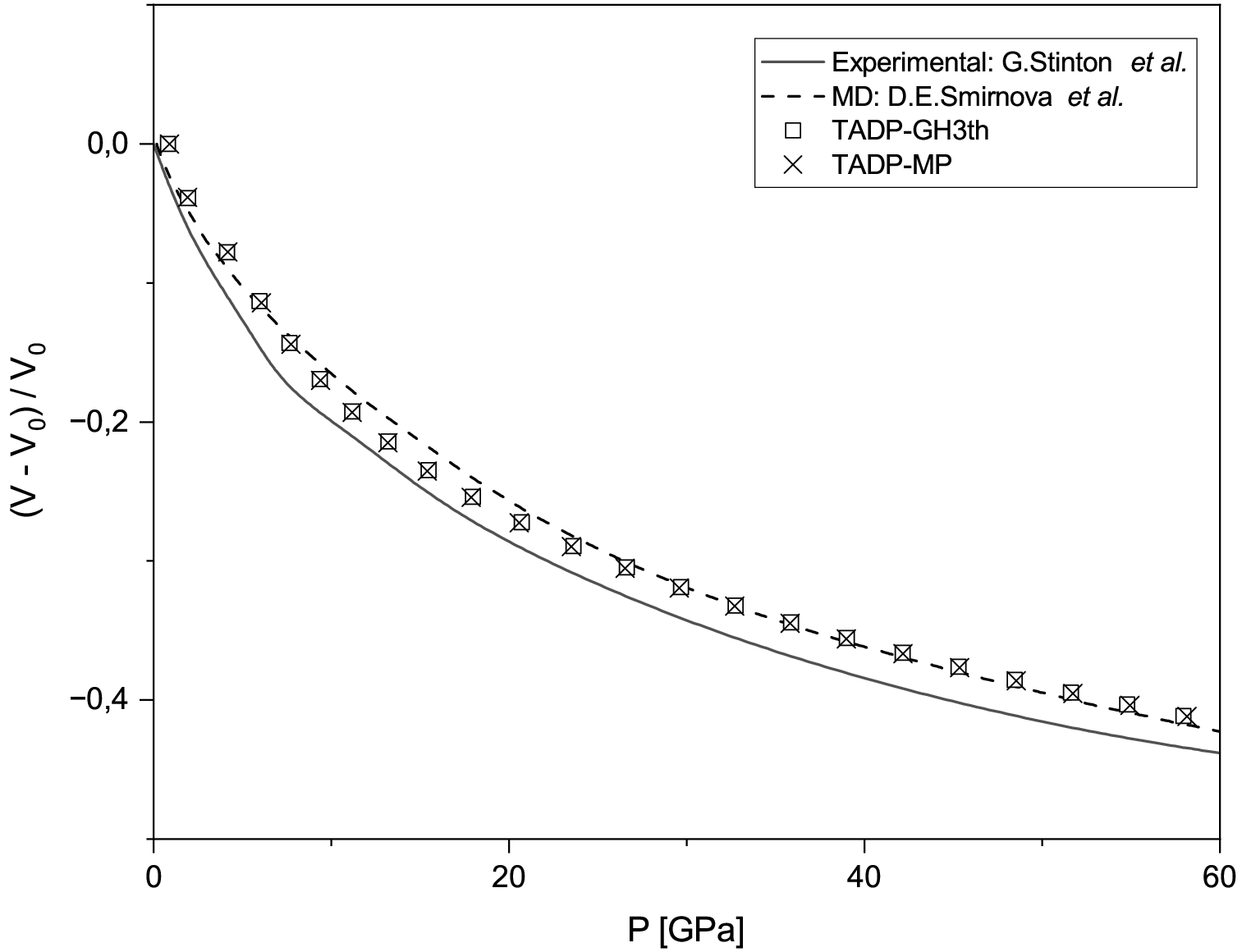}
    \caption{HCP Mg. Calculated isothermal compression curve $T = \SI{300}{\kelvin}$. Molecular dynamics calculations of \cite{SMIRNOVA} and experimental data from \cite{Stinton_et_al_2014} are also shown for comparison.}
    \label{fig:compression-curve}
\end{figure}

As a test of the predicted volumetric equation of state, we calculate the isothermal compression curve of HCP Mg at T = $\SI{300}{\kelvin}$ up to $60$GPa. The results of the calculations are shown in Fig.~\ref{fig:compression-curve}, together with MD calculations of Smirnova~\cite{SMIRNOVA} and experimental data from \cite{Stinton_et_al_2014} by way of comparison. As is evident from the figure, the agreement between the three isothermal compression curves is excellent over the entire pressure range.

\subsection{Elastic constants}

\begin{figure}
    \centering
    \includegraphics[width=0.6\columnwidth]{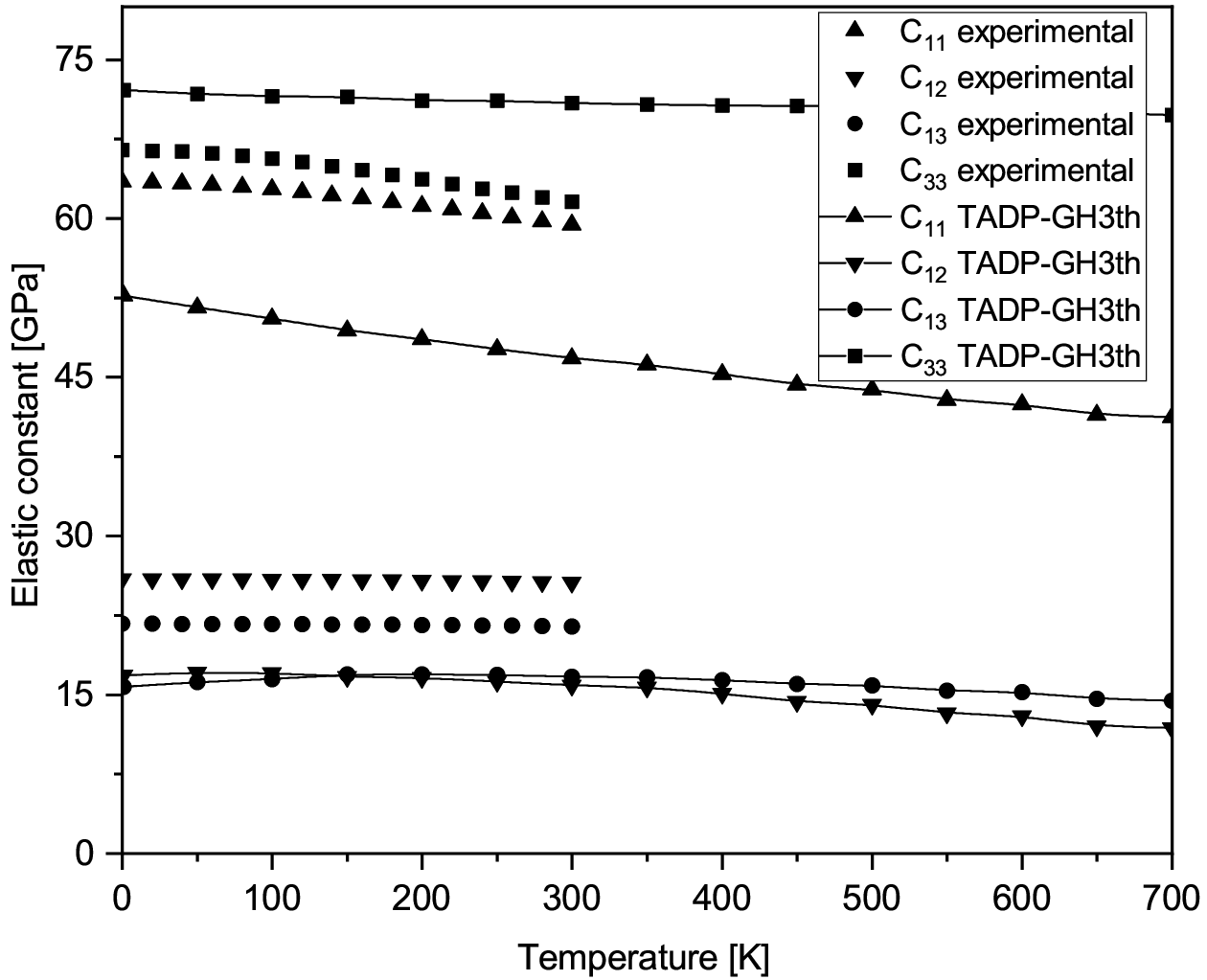}
    \caption{HCP Mg. Calculated dependence of elastic moduli on temperature at zero pressure using second-order multipole quadrature (MP) and third-order Gauss-Hermite quadrature (GH3th). Experimental data of \cite{Slutsky_and_Garland} are also shown for comparison.}
    \label{fig:Cij-curve-Mg}
\end{figure}

\begin{figure}
    \centering
   \includegraphics[width=0.6\columnwidth]{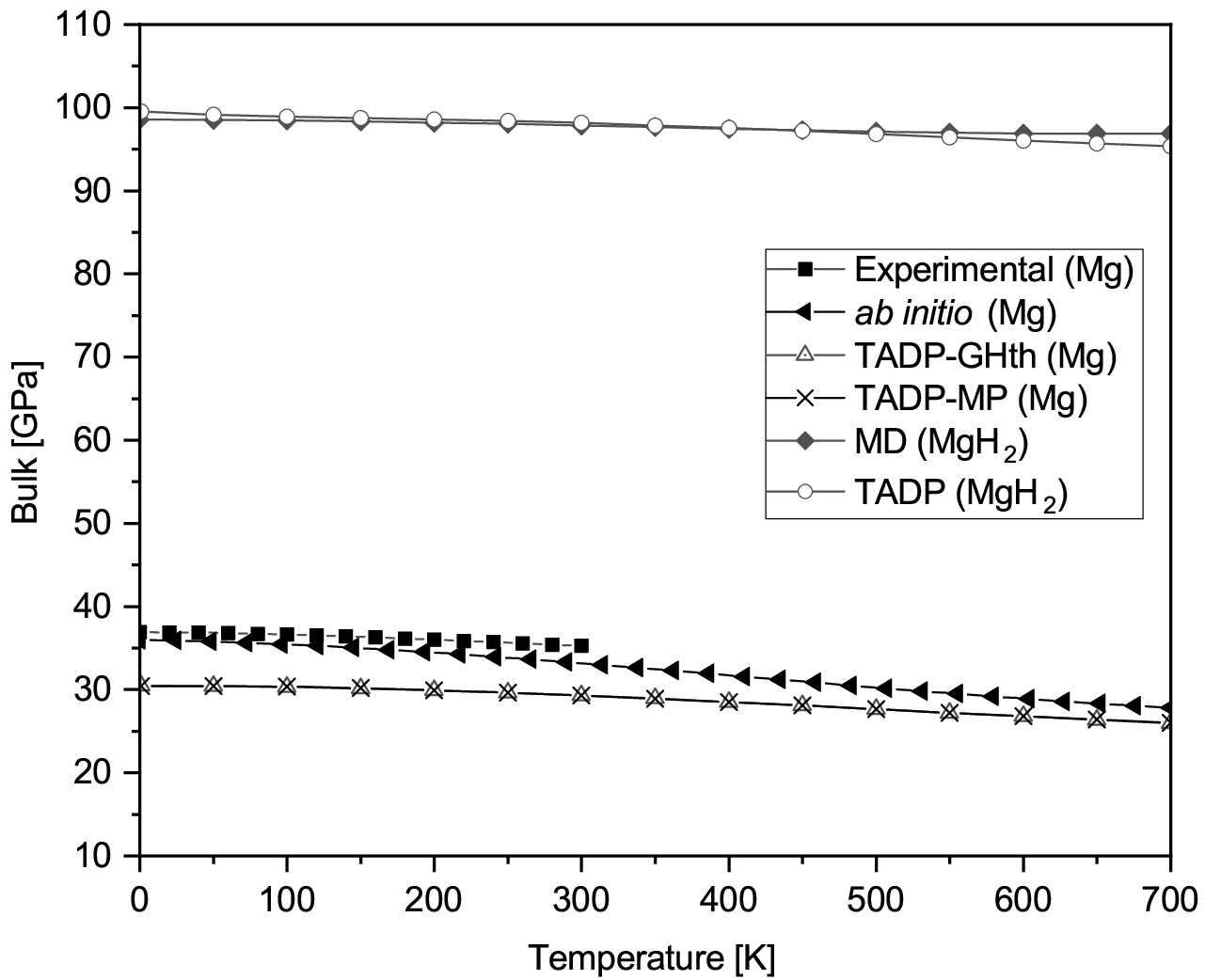}
    \caption{HCP Mg. Calculated dependence of the bulk elastic modulus on temperature at zero pressure using second-order multipole quadrature (MP) and third-order Gauss-Hermite quadrature (GH3th). {\sl ab initio} calculations of \cite{Junkaew_et_al_2014} and experimental data of \cite{Slutsky_and_Garland} are also shown for comparison.}
    \label{fig:Bulk-curve-Mg}
\end{figure}

The calculated temperature dependence of the bulk modulus and elastic constants for HCP Mg and Rutile MgH$_2$ is shown in Figs.~\ref{fig:Cij-curve-Mg} and \ref{fig:Bulk-curve-Mg}, respectively. Additionally, 0K$^{\circ}$ results are tabulated in Tables~\ref{tab:elastic} and {\ref{tab:elastic-MgH2}} for comparison against MD calculations using ADP \cite{SMIRNOVA} and EAM \cite{Sun_et_al_2006}, {\sl ab initio} calculations \cite{Junkaew_et_al_2014} and experimental data \cite{Slutsky_and_Garland}. We observe that the method of thermalization is able to capture the temperature dependence of the elastic moduli predicted by MD. At 0K$^{\circ}$ TADP and MD \cite{SMIRNOVA} show almost identical values, as required, but differ from other potentials and from experimental data. These discrepancies are largest for MgH$_2$, where ADP fails to predict realistic values of the bulk modulus, and identify fidelity limitations of ADP.

\begin{table}
    \centering
    \begin{tabular}{cccccc}
    \hline
         (GPa)     & $C_{11}$     & $C_{33}$     & $C_{12}$     & $C_{13}$  & $B$   \\
    \hline
    Experiments    & 63.5    & 66.5    & 26.0    & 21.7 & 36.9  \\
    MD-ADP   & 52      & 74      & 18      & 16  &  30.88 \\
    MD-EAM & 69.6 & 69.5 & 25.3 & 16 & 35.9 \\
    {\it ab initio} & 68.6 & 73.4 & 23.6 & 17.8 & 36.6 \\
    TADP & 53.4 & 74 & 16.8 & 15.8 & 30.84 \\
    \hline
    \end{tabular}
    \caption{{HCP Mg. Elastic moduli and bulk modulus at zero pressure and zero temperature. TADP compared with MD calculations using ADP \cite{SMIRNOVA} and EAM \cite{Sun_et_al_2006}, {\sl ab initio} calculations \cite{Junkaew_et_al_2014} and experimental data \cite{Slutsky_and_Garland}}. The thermal evolution of the elastic variables is presented in Figs.~\ref{fig:Bulk-curve-Mg} and \ref{fig:Cij-curve-Mg}.}
    \label{tab:elastic}
\end{table}

\begin{table}[]
    \centering
    \begin{tabular}{cccccc}
    \hline
              & $C_{11}$     & $C_{33}$     & $C_{12}$     & $C_{13}$  & $B$   \\
    \hline
    MD-ADP & 131.8 & 123.6 & 90.8 & 69.9 & 98.5 \\
    {\it ab initio} & 74.4 & 136.0 & 38.8 & 31.4 & 54.2 \\
    TADP & 135.6 & 164.7 & 93.4 & 66.9 & 98.9 \\
    \hline
    \end{tabular}
    \caption{{Rutile MgH${}_2$. Elastic moduli and bulk modulus at zero pressure and zero temperature. TADP compared with MD calculations using ADP \cite{SMIRNOVA} and {\sl ab initio} calculations \cite{Junkaew_et_al_2014}}. The thermal evolution of the MgH${}_2$ bulk modulus (B) is presented in Fig.~\ref{fig:Bulk-curve-Mg}.}
    \label{tab:elastic-MgH2}
\end{table}

\section{Summary and concluding remarks}\label{conclusions}

In this study, we have applied a number of computationally efficient approximation techniques, including meanfield approximation and numerical quadrature, to the evaluation of Angular-Dependent interatomic Potentials (ADPs), such as proposed by \cite{SMIRNOVA} for magnesium and magnesium hydrides, at finite temperature (thermalization) and arbitrary atomic molar fractions (mixing). The approach characterizes local equilibrium properties \cite{Kulkarni2006, kulkarni2008, Venturini2011, Venturini2014}  efficiently, in the sense of computational time, and reliably, in the sense of comparable accuracy to Molecular Dynamics (MD). We have numerically verified and experimentally validated the accuracy and fidelity of the resulting thermalized/mixed ADPs (TADPs) by means of selected numerical tests including free entropy, heat capacity, thermal expansion, molar volumes, equation of state and elastic constants. In particular, we have shown that the local equilibrium properties predicted by TADPs agree closely with those computed directly from ADP by means of MD.

The resulting thermalized and mixed ADPs can be used {\sl on-the-fly} as a basis for Diffusive Molecular Dynamics (DMD) simulations of long-term mass transport in magnesium hydrides, including phase transitions and coupling to hydrogen diffusion thereof. This computational paradigm has been successfully applied to other materials systems, such as Pd hydrides \cite{Wang:2015, Sun:2017, sun2018, sun2019}, in order to understand the coupling between hydrogen diffusion and phase boundary structure and mobility. The Mg-H system is particularly challenging because of the displacive HCP-rutile phase transitions undergone upon hydration, which result in complex domain and boundary structures that are closely coupled to hydrogen transport and which compromise the storage capacity and structural integrity of the system \cite{li2007magnesium}. A full characterization of hydride systems thus requires the predictive simulation of strongly coupled mechanical and diffusive phenomena spanning atomistic and macroscopic time scales, a computational challenge well beyond MD but accessible to DMD. These and other extensions of the present work suggest themselves as worthwhile areas of future research.

\section*{Acknowledgements}

M.P.~Ariza gratefully acknowledges financial support from the Ministerio de Ciencia e Innovación under grant number PID2021-124869NB-I00. M.~Molinos gratefully acknowledges the support of the Ministerio de Ciencia e Innovación of Spain for his postdoctoral fellowship under reference FJC2021-046501-I. M.~Ortiz gratefully acknowledges the support of the Deutsche Forschungsgemeinschaft (DFG, German Research Foundation) {\sl via} project 211504053 - SFB 1060; project 441211072 - SPP 2256; and project 390685813 -  GZ 2047/1 - HCM.

\begin{appendix}

\section{Hermitian quadrature}
\label{Hermitian-quadrature}

By an appropriate change of variables, the integrals of interest can be reduced to the standard form
\begin{equation}\label{Integral}
\begin{split}
    &
    I(f) =
    \int_{-\infty}^\infty
    \cdots
    \int_{-\infty}^\infty
    f(x_1,\dots,x_n)
    \times \\ & \qquad\qquad\qquad
    \exp[-x_1^2 - \cdots - x_n^2]
    dx_1\dots dx_n .
\end{split}
\end{equation}
For instance,
\begin{equation}
\begin{split}
&    \meanfield{V} \approx \int^{\infty}_{-\infty}\ V(\{{q}\} , \{\chi\}) \\
&   \prod_{j\in{I_\text{M} \cup  I_\text{H}}}^{N} \frac{1}{(\sqrt{2\pi}\bar{\sigma}_j)^3} \exp \Big(- \frac{1}{2\bar{\sigma}_j^2}|{q}_j - \bar{{q}}_j|^2 \Big) dV_q = \\
    & \int^{\infty}_{-\infty} V(\{{\bar{{q}} + \bar{\sigma}\sqrt{2}\zeta}\}, \{{\chi}\}) \prod_{j\in{I_\text{M} \cup  I_\text{H}}}^{N} \frac{1}{\pi^{n/2}} \text{exp}\left(- |\zeta_j|^2\right) dV_{\zeta}= \\
    & \frac{1}{\pi^{n N/2}} \int^{\infty}_{-\infty}\ V(\{{\bar{{q}} + \bar{\sigma}\sqrt{2}\zeta}\}, \{{\chi}\}) \prod_{j\in{I_\text{M} \cup  I_\text{H}}}^{N} \text{exp}\left(- |\zeta_j|^2\right) dV_{\zeta} \approx \\
    & \frac{1}{\pi^{n N/2}} \sum_{k=1}^{nM} V(\{{\bar{{q}} + \bar{\sigma}\sqrt{2}\zeta^k}\}, \{{\chi}\})\ W^k ,
\end{split}
\end{equation}
which is in the standard form (\ref{Integral}).

An $M$-point numerical quadrature approximates the integral as the sum
\begin{equation}\label{Approximation}
    I(f) \approx \sum_{k=1}^M f(\nu_k)W_k \,,
\end{equation}
where $\nu_k$ is an $n$-dimensional vector at the $k^{th}$ quadrature point
\begin{equation}
    \mathbf{\nu}_k = {\nu_{1k},\ldots, \nu_{nk}} \,.
\end{equation}
Expressions and tables for quadrature rules of various orders and dimensions are presented in \cite{Stroud}.

The third-degree quadrature uses $2n$ points. The points and coefficients are obtained by requiring that the formula integrate all monomials of degree $\leq 3$ exactly. Since the domain of integration is all of $\mathbb{R}^n$ and the Gaussian weight has the property
$w(x) = w(-x)$, the distribution of quadrature points is fully symmetric, i.~e., the coefficient of $\nu_k$ equals that of $-\nu_k$. Thus, for an $n$-dimensional space, the points are
\begin{equation}
     (\pm\, r,0, \ldots, 0) \quad \dots \quad (0,\ldots, 0, \pm\, r),
\end{equation}
with coefficients
\begin{equation}
    W_k = \frac{V}{2n},
    \qquad
    k = 1, \ldots, n \,,
\end{equation}
where
\begin{equation}
    V = I(1) = \pi^{n/2},
    \qquad
    r^2 = \frac{n}{2}\,.
\end{equation}

\section{Multipole quadrature}
\label{Multipole-quadrature}

Let $\mu$ be a bounded measure over $\mathbb{R}^n$ and let $f \in C(\mathbb{R}^n)$ be a continuous function. We write the action of the measure $\mu$ on $f$ as
\begin{equation}
    \mu(f) = \int f(x) \, d\mu(x) .
\end{equation}
Then, the multipole approximation of $\mu$ of degree $k$ is the measure $\mu_k$  such that
\begin{equation}
    \mu_k(f)
    =
    \sum_{|\alpha|\leq k}
    c_\alpha D^\alpha f(0)
\end{equation}
where $\alpha$ is a multiindex in $\mathbb{N}^n$ and the coefficients $\{c_\alpha,\ |\alpha| \leq k\}$ of the approximation are chosen such that the approximation is exact for polynomials of degree $\leq k$, i.~e., such that
\begin{equation}
  \label{eq-multipole-def}
    \mu_k(x^\alpha) = \mu(x^\alpha),
    \qquad
    |\alpha| \leq k .
\end{equation}
This requirement gives
\begin{equation}
    c_\alpha
    =
    \frac{1}{\alpha!} \int x^\alpha \, d\mu(x) .
\end{equation}
For one-dimensional Gaussian integrals, i.~e., for the measure
\begin{equation}\label{eq:ot1umo}
    d\mu(x)
    =
    \sqrt{\frac{a}{2\pi}} {\rm e}^{-a \frac{x^2}{2}}
    \, dx ,
\end{equation}
the corresponding coefficients can be computed explicitly, with the result,
\begin{equation}
\begin{split}
   & c_m
    =
    \frac{2^{n/2} a^{-n/2} \Gamma \left(\frac{n+1}{2} \right)}{\sqrt{\pi } n!} ,
    \quad
    n \text{ even} ;\\
  &  c_m = 0,
    \qquad \qquad \qquad \qquad \quad
    n \text{ odd} .
\end{split}
\end{equation}
The first five non-zero coefficients evaluate to
\begin{equation}
    c_0 = 1 , \;
    c_2 = \frac{1}{2a} , \;
    c_4 = \frac{1}{8a^2} , \;
    c_6 = \frac{1}{48a^3} , \;
    c_8 = \frac{1}{384a^4} .
\end{equation}
We note that, with the exception of $c_0$, these coefficients diverge when $a \downarrow 0$.

\end{appendix}

\bibliographystyle{unsrt}

\end{document}